\documentclass[10pt,twoside]{article} 
\usepackage{graphicx}
\usepackage{verbatim}
\usepackage{framed}
\usepackage{hyperref}
\usepackage{amsmath}
\usepackage[normalem]{ulem}

\begin{document}

\title{Efficient depth extrapolation of waves in elastic isotropic media}
\author{Musa Maharramov (\href{mailto:maharram@stanford.edu}{maharram@stanford.edu})}

\date{May 08, 2012 rev 1.0 \\ May 19, 2012 rev 2.0}

\maketitle

\begin{abstract}
We propose a computationally efficient technique for extrapolating seismic waves in an arbitrary isotropic elastic medium. The method is based on factorizing the full elastic wave equation into a product of pseudo-differential operators. The method extrapolates displacement fields, hence can be used for modeling both pressure and shear waves. The proposed method can achieve a significant reduction in the cost of elastic modeling compared to the currently prevalent time- and frequency-domain numeric modeling methods and can contribute to making multicomponent elastic modeling part of the standard seismic processing work flow.
\end{abstract}

\setcounter{secnumdepth}{2}

\section{Introduction}

Extrapolation of seismic wave fields in depth using one-way propagation operators is an efficient alternative to time- and frequency-domain modeling with the full wave equation, particularly in seismic migration applications (see  \cite{IEI}, \cite{Biondo}). While one-way extrapolators have long been established as key components of the seismic imaging toolbox for isotropic acoustic media, extrapolation of elastic wave fields is still carried out by solving the full elastodynamic system either in the time or frequency domain, either approach being computationally expensive. The high computational cost of wave extrapolation in elastic media is one of the barriers to a widespread adoption of multicomponent seismic in industrial applications. Some progress has been made recently in the development of efficient one-way methods for certain simplest \emph{anisotropic} elastic models (e.g., vertically transversally isotropic or tilted transversally isotropic media -- see \cite{SHAN07}, \cite{NOLTE08}, \cite{MUSAEAGE11}) However, these methods use the ``pseudoacoustic'' approximation (see \cite{GRECHKA}) and are used for a kinematically accurate propagation of pressure waves only.

In this paper we present a method for one-way frequency-domain extrapolation of \emph{displacement} fields in an elastic isotropic medium. The approach of this paper is based on factorizing elastic wave equation using pseudo-differential operators without introducing stress-related unknown functions into the equations. Our approach is conceptually similar to the derivation of the acoustic single square-root equation (see \cite{IEI}) except the resulting factorized propagation operators can not be obtained analytically but are computed numerically.

\section{The Method}

We start with the wave equation governing the displacement in an arbitrary heterogenous isotropic elastic medium in the \emph{Navier form} (see \cite{SEGDEF}):
\begin{equation}
\rho \ddot{u^i}\;=\;\mu \Delta u^i+ \frac{\mu}{1-2\nu} \frac{ \partial}{\partial x_i} \frac{ \partial u_k}{\partial x_k},\;i=1,2,3,
\label{eq:ewe}
\end{equation}
where $u^i$ denote the components of a displacement field, $\mu$ is the shear modulus, $\nu$ is Poisson's ratio for the medium, and $\rho$ is the density. In this paper we consider a heterogenous elastic medium under the assumption of \emph{local homogeneity} -- otherwise the elastic moduli would not be factored outside of the differentiation operators in equation \ref{eq:ewe}. However, our method can be extended to cover the case when local homogeneity assumption is dropped. ``Freezing'' the coefficients of equation \ref{eq:ewe} and applying the Fourier transform in time and horizontal variables $x_1=x,x_2=y$, and substituting
\begin{equation}
\frac{\mu}{1-2\nu}=\lambda+\mu,
\label{eq:lam}
\end{equation}
where $\lambda$ is the Lam\'{e} coefficient (see \cite{RPH},\cite{SEGDEF}), we get
\begin{align}
\rho \omega^2 u^1+\mu\left[ (-k_x^2-k_y^2)u^1+\frac{\partial^2 u^1}{\partial z^2}\right]+(\lambda+\mu)\left[ -k_x^2 u^1-k_x k_y u^2+ik_x\frac{\partial u^3}{\partial z}\right] =& 0, \nonumber \\
\rho \omega^2 u^2+\mu\left[ (-k_x^2-k_y^2)u^2+\frac{\partial^2 u^2}{\partial z^2}\right]+(\lambda+\mu)\left[ -k_x k_y u^1-k_y^2 u^2+ik_y\frac{\partial u^3}{\partial z}\right] =& 0, \nonumber \\
\rho \omega^2 u^3+\mu\left[ (-k_x^2-k_y^2)u^3+\frac{\partial^2 u^3}{\partial z^2}\right]+(\lambda+\mu)\left[ i k_x \frac{\partial u^1}{\partial z}+i k_y \frac{\partial u^2}{\partial z} + \frac{\partial^2 u^3}{\partial z^2}\right] =& 0,
\label{eq:fewe}
\end{align}
where $k_x,k_y$ are horizontal wave numbers and $\omega$ is the frequency. The left-hand side of system \ref{eq:fewe} is the result of an ordinary differential operator applied to a vector-function $\mathbf{u}=(u^1,u^2,u^3)$ and parametrized by horizontal wave numbers. In the present form equations \ref{eq:fewe} cannot be used for computationally efficient explicit depth extrapolation in a heterogeneous medium; however, these equations can be used for modeling displacements by solving a series of boundary-value problems (see \cite{MUSASEP147}). In \cite{MUSASEP147} it was suggested that equations \ref{eq:fewe} might be factorized in such a way as to allow solving them by alternating \emph{one-way extrapolation} in opposite directions. More specifically, we seek a factorization of operator equation \ref{eq:fewe} of the form:
\begin{equation}
\left(E(\lambda,\mu) \frac{\partial}{\partial z} +A(k_x,k_y)+c_{\omega} I \right) \times
\left(E(\lambda,\mu) \frac{\partial}{\partial z} +B(k_x,k_y)+c_{\omega} I \right) \mathbf{u}=0,
\label{eq:fac}
\end{equation}
where
\begin{align}
E(\lambda,\mu)\; & = \;\left[ \begin{array}{ccc}
\sqrt{\mu} & 0 & 0 \\
0 & \sqrt{\mu} & 0 \\
0 & 0 & \sqrt{\lambda+2\mu} \end{array} \right], \nonumber \\
c_{\omega} \; & = \; \sqrt{\rho}\omega,
\end{align}
and $A,B$ are $3\times 3$ matrices with components that are complex-valued functions of the horizontal wave numbers, $I$ is the $3\times 3$ identity matrix. Performing the multiplication in equation \ref{eq:fac} and using equation \ref{eq:fewe}, we obtain:
\begin{align}
A(k_x,k_y)B(k_x,k_y)+c_{\omega}[A(k_x,k_y)+B(k_x,k_y)] & =P(k_x,k_y), \nonumber \\
A(k_x,k_y) E(\lambda,\mu)+E(\lambda,\mu) B(k_x,k_y)+2c_{\omega} E(\lambda,\mu) & =S(k_x,k_y),
\label{eq:fac2}
\end{align}
where
\begin{align}
P= & \left[ \begin{array}{ccc}
-(\lambda+2\mu)k_x^2 - \mu k_y^2 & -(\lambda+\mu) k_x k_y & 0 \\
 -(\lambda+\mu) k_x k_y & -(\lambda+2\mu)k_y^2 - \mu k_x^2 & 0 \\
0 & 0 & -\mu(k_x^2 + k_y^2) \end{array} \right], \nonumber \\
S= & \left[ \begin{array}{ccc}
0 & 0 & i (\lambda+\mu) k_x \\
0 & 0 & i (\lambda+\mu) k_y \\
i(\lambda+\mu) k_x & i(\lambda+\mu) k_y & 0 \end{array} \right].
\label{eq:ps}
\end{align}
Combining equations \ref{eq:fac2} and \ref{eq:ps}, we get the following equation for the operators $A$ and $B$:
\begin{align}
A(k_x,k_y)B(k_x,k_y)+c_{\omega}(A(k_x,k_y)+B(k_x,k_y))=P(k_x,k_y), \nonumber \\
E(\lambda,\mu) B(k_x,k_y)+A(k_x,k_y) E(\lambda,\mu)=\tilde{S}(k_x,k_y),
\label{eq:main}
\end{align}
where
\begin{equation}
\tilde{S}(k_x,k_y)= S(k_x,k_y)-2 c_{\omega} E(\lambda,\mu).
\label{eq:ps2}
\end{equation}
Equations \ref{eq:fac}, \ref{eq:main} in combination with equations \ref{eq:ps} and \ref{eq:ps2} suggest the following procedure for extrapolating solutions to system \ref{eq:ewe} in depth:
\begin{enumerate}
\item{Solve the system of matrix equations \ref{eq:main} for $A,B$, for each pair of horizontal wave numbers $k_x,k_y$ and two reference values of each elastic parameter $\lambda_{\textrm{min}},\lambda_{\textrm{max}}$ and  $\mu_{\textrm{min}},\lambda_{\textrm{max}}$;
}
\item{Evaluate
\[\left( E(\lambda,\mu) \frac{\partial}{\partial z}+B(-i\partial_x, -i\partial_y)+ c_{\omega} I \right) \mathbf{u}(x,y,z=0)\]
from the initial conditions and assign the value to an auxiliary function $\tilde{\mathbf{u}}(x,y,z=0)$;
}
\item{Solve 
\begin{equation}
\left( E(\lambda,\mu) \frac{\partial}{\partial z}+A(-i\partial_x, -i\partial_y)+ c_{\omega} I \right) \mathbf{\tilde{u}}(x,y,z)=0
\label{eq:down}
\end{equation}
by downward continuing in depth, using the formula
\begin{equation}
\tilde{u}(x,y,z+\Delta z)=\exp{\left[ -\Delta z E^{-1}(A(-i\partial_x,-i \partial_y)+c_{\omega}I)\right]}\mathbf{\tilde{u}}(x,y,z).
\label{eq:downexp}
\end{equation}
}
\item{Perform each step of the depth extrapolation for four combinations of the reference elastic parameters, then apply inverse Fourier transform to the four fields and interpolate at each spatial point of the depth slice using true $\lambda(x,y),\mu(x,y)$ as e.g. in the PSPI method (see \cite{Biondo}).}
\item{After reaching the desired maximum depth, find the solution $\mathbf{u}$ by \emph{upward} extrapolation:
\begin{equation}
\left( E(\lambda,\mu) \frac{\partial}{\partial z} +B(-i\partial_x, -i\partial_y)+ c_{\omega} I \right) \mathbf{u}(x,y,z)= \mathbf{\tilde{u}}(x,y,z).
\label{eq:up}
\end{equation}
}
\item{Repeat the above steps for each frequency component $\mathbf{u}(\omega,x,y,z).$}
\end{enumerate}
The above algorithm is stable if the spectrum of matrix
\begin{equation}
A(k_x,k_y)+c_{\omega} I
\label{eq:downmat}
\end{equation}
is not in the interior of the left half-plane, and the spectrum of
\begin{equation}
B(k_x,k_y)+c_{\omega} I
\label{eq:upmat}
\end{equation}
is not in the interior of the right half-plane.
While the above algorithm tries to mimic the two-way wave propagation, it is effectively just an approximation to the propagation process as it ignores the interaction between the up and down-going wave at intermediate depth steps. A less accurate alternative would be to downward-continue the wave field using equation \ref{eq:downexp} in a way similar to the one-way depth extrapolation using the scalar square-root equation (see \cite{IEI},\cite{Biondo}). The latter approach would be unable to image any dips beyond 90$^\circ$, however, it would reduce the cost of extrapolation by a further factor of 2. Note the cost of solving equation \ref{eq:down} in depth is roughly three times that of solving the scalar square-root equation.

The above analysis may be extended to the case of an arbitrary anisotropic elastic medium. The fact that the components of the pseudo-differential operator matrices $A(-i\partial_x,-i\partial_y), B(-i\partial_x,-i\partial_y)$ are not given in an analytical form but are only computed numerically does not limit their applicability.

Factorization of system \ref{eq:fewe} in the elastostatic case was one of the approaches mentioned by the author in \cite{MUSASEP147}. However, the one-way extrapolation technique is mostly useful for elastodynamic problems as the passband of the factorized depth extrapolators (e.g., as in equation \ref{eq:downexp}) narrows down to zero with the temporal frequency passing to the zero static limit. 
\begin{figure}[htb]
\begin{center}
\includegraphics[width=1.\textwidth]{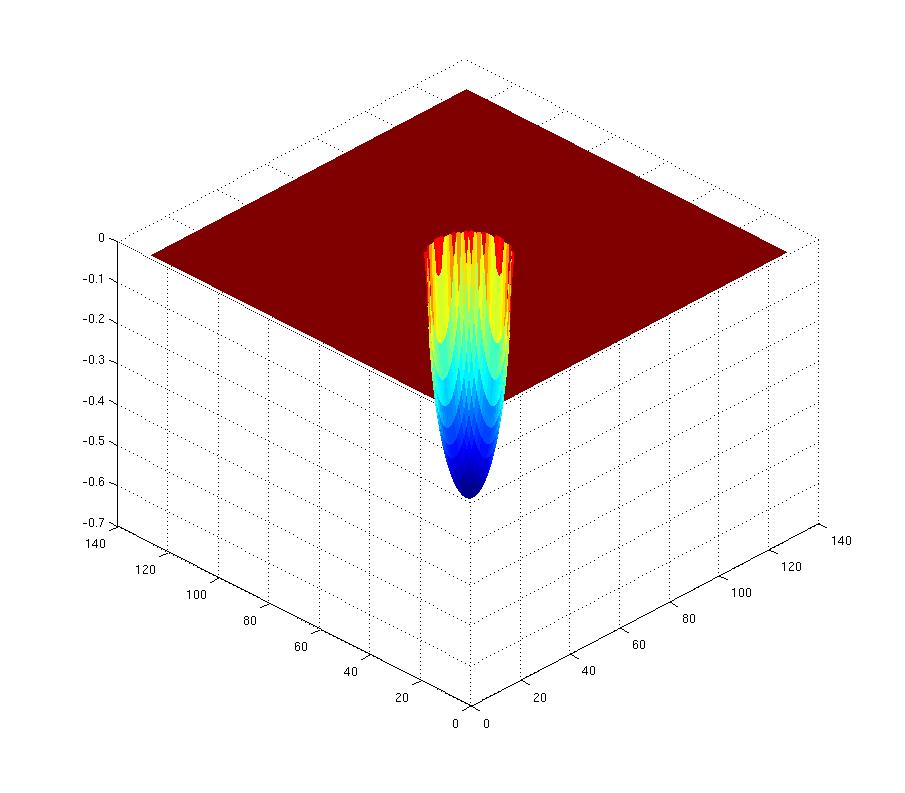}
\end{center}
\caption{The phase of a phase-shift operator corresponding to the maximum imaginary part of the eigenvalues of operator \ref{eq:logextrap}. Multicomponent ``phase-shift'' is defined by three such scalar phase-shift operators and a $3\times 3$ matrix $Q$ of equation \ref{eq:Kfact}.}
\label{fig:maximageigenval}
\end{figure}
Note that equation \ref{eq:ewe} uses elastic parameterization that degenerates into a singularity if the shear modulus is equal to zero. This is not causing any problems with purely acoustic wave extrapolation as the singularity is effectively removed from equations \ref{eq:fewe} by the substitution \ref{eq:lam}.

\section{Implementation and Results}

\begin{figure}[htb]
\begin{center}
\includegraphics[width=.8\textwidth]{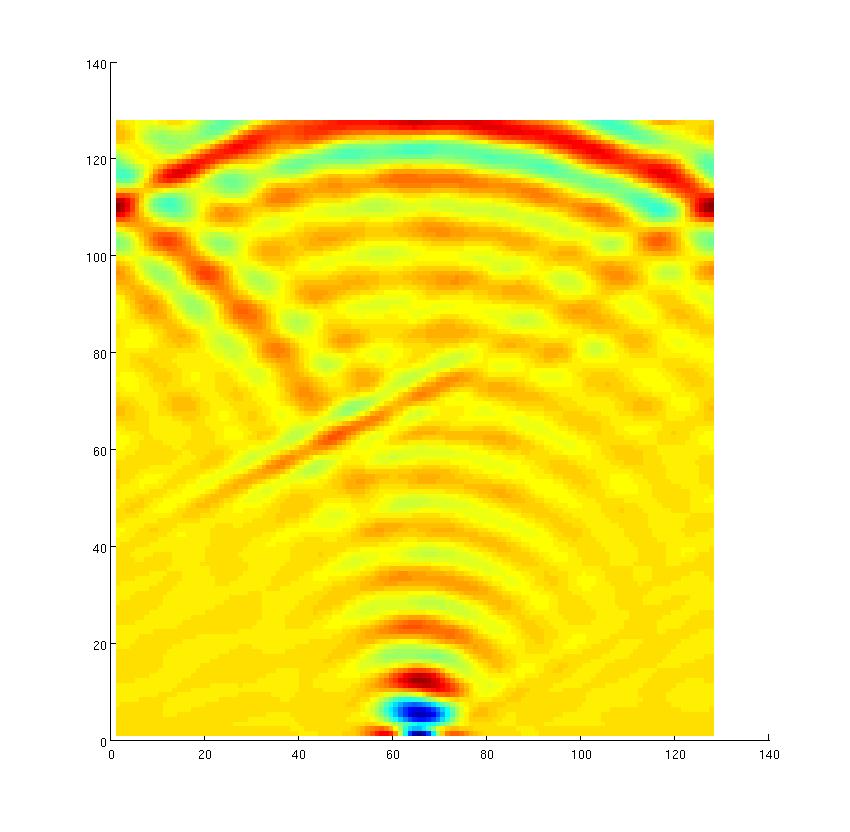}
\end{center}
\caption{Pressure wave extrapolated from an impulse displacement, 2-12 \texttt{Hz}, $128\times 128 \times 128$ grid, $5$ \texttt{m} step, $316$ \texttt{m/s} shear-wave and $632$ \texttt{m/s} pressure-wave velocity.}
\label{fig:pressure}
\end{figure}

\begin{figure}[htb]
\begin{center}
\includegraphics[width=.8\textwidth]{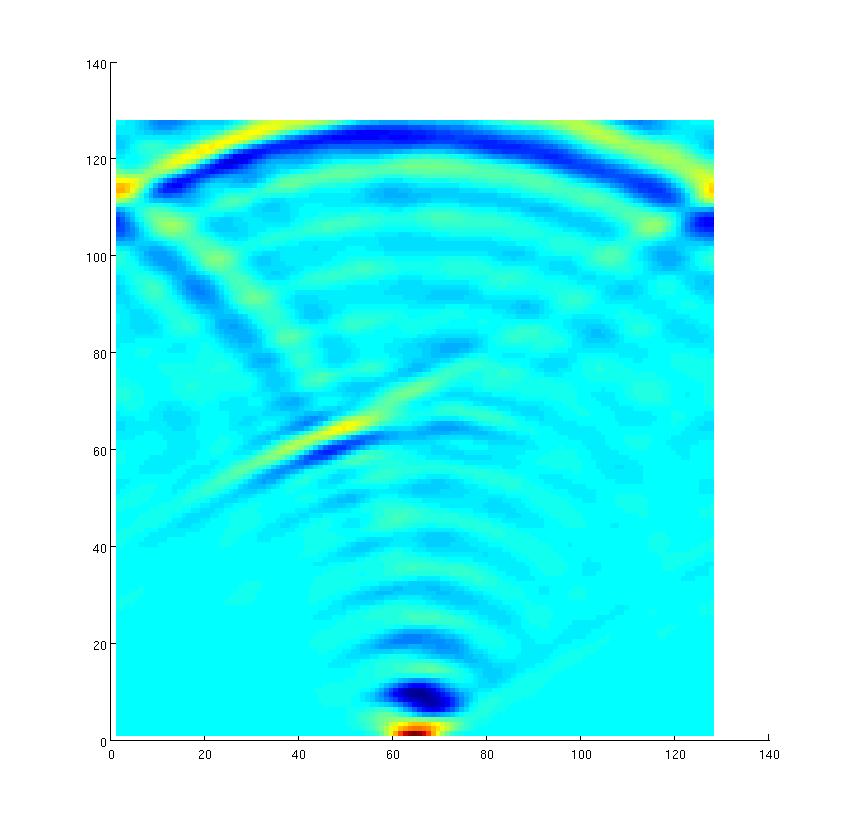}
\end{center}
\caption{Vertical component of a wave extrapolated from an impulse displacement, inline section, 2-12 \texttt{Hz}, $128\times 128$ grid, $5$ \texttt{m} step, $316$ \texttt{m/s} shear-wave and $632$ \texttt{m/s} pressure-wave velocity.}
\label{fig:component3}
\end{figure}

\begin{figure}[htb]
\begin{center}
\includegraphics[width=.8\textwidth]{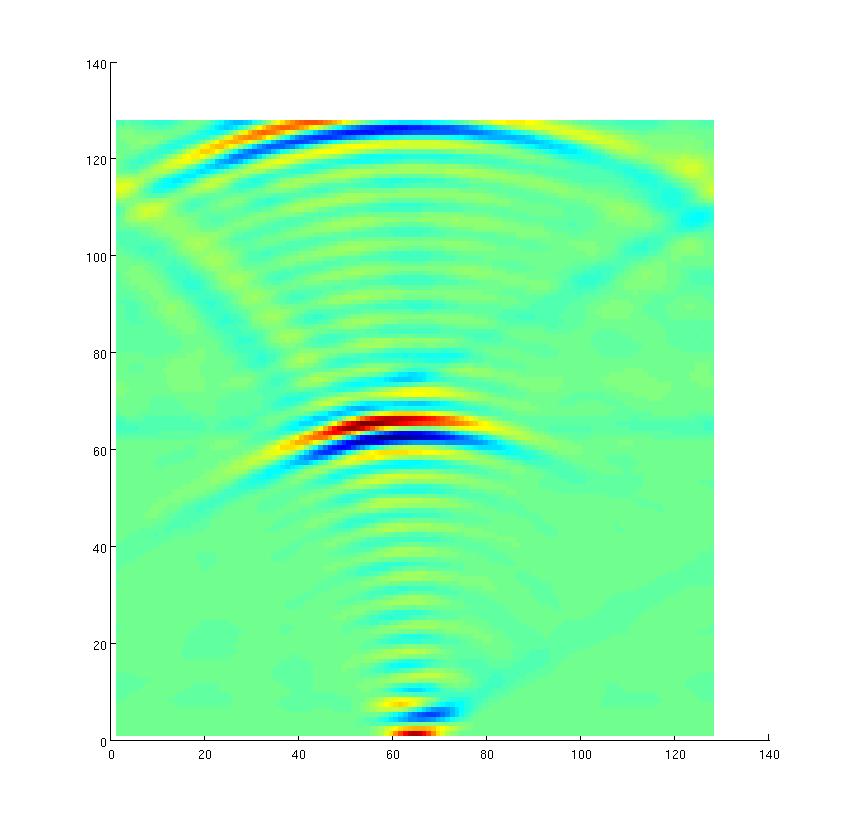}
\end{center}
\caption{Inline component of a wave extrapolated from an impulse displacement, inline section, 2-12 \texttt{Hz}, $128\times 128 \times 128$ grid, $5$ \texttt{m} step, $316$ \texttt{m/s} shear-wave and $632$ \texttt{m/s} pressure-wave velocity. Note the slow shear wave.}
\label{fig:component1}
\end{figure}

\begin{figure}[htb]
\begin{center}
\includegraphics[width=.8\textwidth]{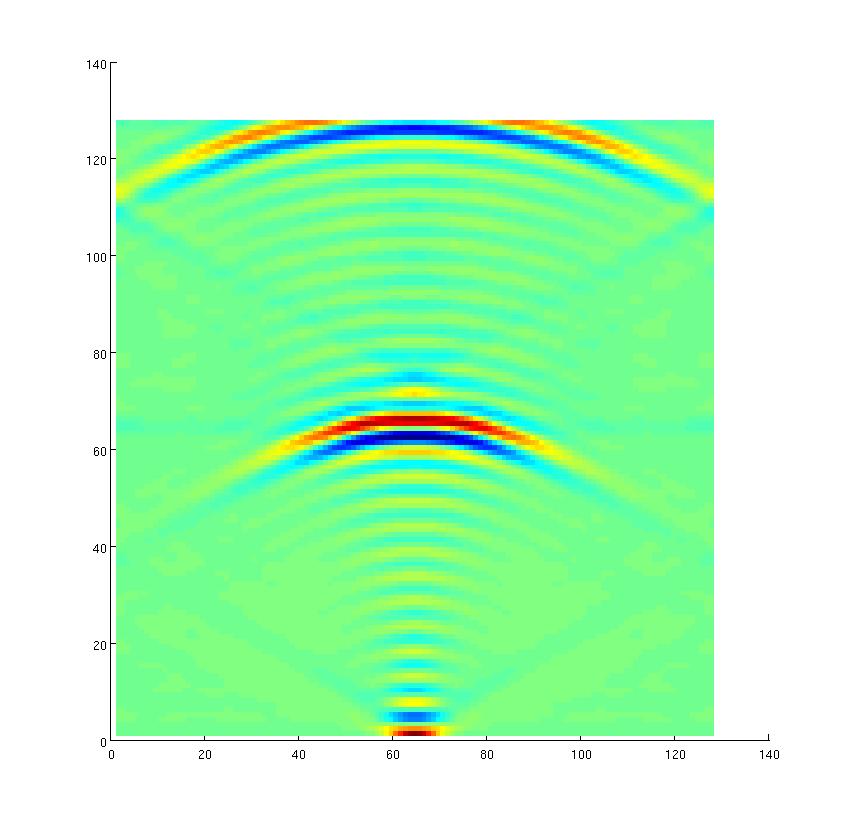}
\end{center}
\caption{Crossline component of a wave extrapolated from an impulse displacement, inline section, 2-12 \texttt{Hz}, $128\times 128 \times 128$ grid, $5$ \texttt{m} step, $316$ \texttt{m/s} shear-wave and $632$ \texttt{m/s} pressure-wave velocity. Note the slow shear wave.}
\label{fig:component2}
\end{figure}

The system of matrix equations \ref{eq:main} is solved only once for each triple of temporal frequency and elastic moduli values, and for each pair of horizontal wave numbers. In our prototype implementation of the one-way extrapolator we compute the matrices $A,B$ at the beginning of the frequency loop and subsequently use the tabulated matrices in the depth extrapolation loop (inside the frequency loop).  A more efficient approach can be employed in a production implementation of the extrapolation method: system \ref{eq:main} can be solved using Newton's method (see e.g. \cite{HIGHAM}) in a one-off computation for each set of the temporal frequency, elastic moduli and horizontal wave numbers and stored in a look-up table. The symmetry of the extrapolation operators \ref{eq:downexp} that appear to be multi-component counterparts of the acoustic phase-shift operator (see \cite{IEI}) can be exploited to achieve a substantial reduction in the size of the precomputed operator tables. Figure \ref{fig:maximageigenval} is the plot of the maximum of the imaginary parts of the three eigenvalues of operator
\begin{equation}
K=-\Delta z E^{-1}\left[A(-i\partial_x,-i \partial_y)+c_{\omega}I\right],
\label{eq:logextrap}
\end{equation}
within its passband. The operator is the one used later to produce images of figures \ref{fig:pressure},\ref{fig:component3},\ref{fig:component1},\ref{fig:component2}. The real parts of the eigenvalues of operator \ref{eq:logextrap} are zero within the operator passband and negative outside. The imaginary parts of the other two eigenvalues exhibit similar behavior. Operator $K$ of equation \ref{eq:logextrap} is the logarithm of the extrapolation operator \ref{eq:downexp}, and the spectral plot of figure \ref{fig:maximageigenval} corresponds to the phase of the phase-shift extrapolator in the acoustic case (see \cite{Biondo}). The crucial difference in the elastic multicomponent case is that the multicomponent ``phase-shift'' is defined by three such scalar phase-shift operators with phases $\phi_1,\phi_2,\phi_3$, and a unitary operator $Q$, determined by the eigenvector expansion of $K$ as follows:
\begin{equation}
K=Q\left[ \begin{array}{ccc}
i\phi_1 & 0 & 0 \\
0 & i\phi_2 & 0 \\
0 & 0 & i\phi_3 \end{array} \right]Q^{\ast}.
\label{eq:Kfact}
\end{equation}

The pass bands of the three phase shift operators are, generally, different, but the real parts of the eigenvalues of  \ref{eq:logextrap} are non-positive across all three pass bands.

Figures \ref{fig:pressure},\ref{fig:component3},\ref{fig:component1},\ref{fig:component2} demonstrate the result of applying our method to extrapolating displacement waves from a concentrated impulse at the surface. Medium parameters used in this test were $316$ \texttt{m/s} shear-wave velocity
\[v_{\texttt{S}}=\sqrt{\mu/\rho}\]
 and $632$ \texttt{m/s} pressure-wave velocity
\[v_{\texttt{P}}=\sqrt{(\lambda+2\mu)/\rho}.\]
The extrapolation grid was $128\times 128 \times 128$ with a 5 \texttt{m} step, frequency range 2-12 \texttt{Hz} with 1 \texttt{Hz} step. The values of the elastic parameters used in this test are uncharacteristically low for seismic applications and were chosen solely for the purpose of fast small-scale simulation on a single-core PC using \texttt{Matlab}.


Since the impulse at the surface is an asymmetric horizontal displacement but can be assumed to be symmetric in the vertical direction, our waves are effectively a combination pressure and shear waves for the horizontal components, while the vertical displacement wave should kinematically match the pressure wave. And indeed, the pressure wave plot \ref{fig:pressure} and vertical displacement plot \ref{fig:component3} exhibit excellent kinematic agreement.

The horizontal wave component plots \ref{fig:component1} and \ref{fig:component2}, on the other hand, show pressure- and shear-wave images, both correctly positioned in agreement with the velocity values used in the simulation. Boundary reflections and low frequency content cause some imaging artifacts that are not related to the method.

\section{Conclusions and Discussion}

The method presented in this paper can be used in seismic migration algorithms in order to achieve a substantial reduction of run time in comparison with the reverse time migration. More specifically, stability of the time-domain modeling typically utilized in the reverse-time migration requires time steps significantly smaller than the time resolution of seismic data (see \cite{Biondo}). Depth extarpolation of wave fields using one-way equations \ref{eq:down} and \ref{eq:up} can be performed for an arbitrary frequency range. Extrapolating wave fields in the frequency domain using two-way system \ref{eq:fewe} would require solving a large sparse system of equations using e.g. finite elements method, still posing significant computational challenges for inhomogeneous media. However, the one-way extrapolation method, while limited in dip and less accurate in terms of amplitudes, lends itself to efficient implementation using e.g. PSPI or finite differencing. Furthermore, the approach of this paper can be expected to apply to more complex elastic anisotropic models (see \cite{GRECHKA}) and may be developed into a computationally efficient alternative to the existing pseudo-acoustic anisotropic modeling methods while allowing easy separation of pressure and shear waves.

\section{Acknowledgements}

The author would like to thank the Stanford Exploration Project for supporting this work, and Jon Claerbout, Stewart Levin, Biondo Biondi and Paul Segall for a number of useful discussions.

\bibliographystyle{plain}
\bibliography{musa_maharramov_May08_12.bib}

\begin{thebibliography}{10}

\bibitem{Biondo}
Biondo Biondi.
\newblock {\em 3{D} {Seismic} {I}maging}.
\newblock Society of Exploration Geophysicists, 2005.

\bibitem{IEI}
Jon Claerbout.
\newblock {\em Imaging the earth's interior}.
\newblock Blackwell Scientific, 1985.

\bibitem{GRECHKA}
Vladimir Grechka.
\newblock {\em Applications of Seismic Anisotropy in Oil and Gas Industry}.
\newblock EAGE, 2009.

\bibitem{HIGHAM}
Nicholas Higham.
\newblock {\em Functions of matrices: theory and computation}.
\newblock SIAM, 2008.

\bibitem{MUSASEP147}
Musa Maharramov.
\newblock Identifying reservoir depletion patterns with applications to seismic
  imaging.
\newblock {\em SEP Report 147}, (147):193--218, 2012.

\bibitem{MUSAEAGE11}
Musa Maharramov and Bertram Nolte.
\newblock Efficient one-way wave-equation migration in tilted transversally
  isotropic media.
\newblock {\em 73rd EAGE Conference and Exhibition, Extended Abstracts}, 2011.

\bibitem{RPH}
Gary Mavko, Tapan Mukerji, and Jack Dvorkin.
\newblock {\em The Rock Physics Handbook}.
\newblock Cambridge University Press, 2009.

\bibitem{NOLTE08}
Bertram Nolte.
\newblock Fourier finite-difference depth extrapolation for {VTI} media.
\newblock {\em 70th EAGE Conference and Exhibition, Extended Abstracts}, 2008.

\bibitem{SEGDEF}
Paul Segall.
\newblock {\em Earthquake and Volcano Deformation}.
\newblock Princeton University Press, 2010.

\bibitem{SHAN07}
Guojian Shan.
\newblock Optimized implicit finite-difference migration for {TTI} media.
\newblock {\em 77th SEG Conference and Exhibition, Extended Abstracts}, 2007.

\end{thebibliography}

\end{document}